\documentclass[aps,pre,groupedaddress]{revtex4-2}
\usepackage{natbib}
\usepackage{amsmath}
\usepackage{epsfig}
\usepackage{color}

\usepackage{graphicx}
\usepackage{amsfonts}
\usepackage{dcolumn}
\usepackage{bm}
\usepackage{color}
\usepackage{hyperref}

\begin{document}
\bibliographystyle{unsrt}

\title{Semi-Markov Processes in Open Quantum Systems. III.  Large Deviations of First Passage Time Statistics} 
\author{Fei Liu}
\email[Email address: ]{feiliu@buaa.edu.cn}
\affiliation{School of Physics, Beihang University, Beijing 100083, China}
\author{Shihao Xia, Shanhe Su}
\affiliation{Department of Physics, Xiamen University, Xiamen 361005, China}

\date{\today}

\begin{abstract}
{In a specific class of open quantum systems with finite and fixed numbers of collapsed quantum states, the semi-Markov process method is used to calculate the large deviations of the first passage time statistics. The core formula is an equation of poles, which is also applied in determining the scaled generating functions (SCGFs) of the counting statistics. For simple counting variables, the SCGFs of the first passage time statistics are derived by finding the largest modulus of the roots of this equation with respect to the $z$-transform parameter and then calculating its logarithm. The procedure is analogous to that of solving for the SCGFs of the counting statistics. However, for  current-like variables, the method generally fails unless the equation of pole is simplified to a quadratic form. The fundamental reason for this lies in the nonuniqueness between the roots and the region of convergence for the joint transform. We illustrate these results via a resonantly driven two-level quantum system, where for several counting variables the solutions to the SCGFs of the first passage time are analytically obtained. Furthermore, we apply these functions to investigate quantum violations of the classical kinetic and thermodynamic uncertainty relations.}

\end{abstract}

\maketitle

\section{Introduction} 
\label{section1}
In our preceding three papers~\cite{Liu2022,Liu2023a,Liu2023b}, a semi-Markov process (SMP) method was developed to investigate the large deviations of the counting statistics (LDs-CS)~\cite{Levitov1996,Bagrets2003,Esposito2009,Rudge2019,Landi2024} {for a specific class of open quantum systems. The dynamics of these systems are described by the Markovian quantum master equations (QMEs), and their unraveling of the quantum jump trajectories (QJTs) involves fixed sets of finite collapsed quantum states.} Our results show that the method is not only intuitive in terms of the physical picture but also effective in analyzing and calculating the LDs~\cite{Touchette2008}. Furthermore, when extended to complex situations where random resetting, with or without memory, is involved in quantum processes, the method shows a certain flexibility~\cite{Liu2023b}. When we combine these results with an earlier application of an SMP in deriving a quantum uncertainty relation~\cite{Carollo2019} and the latest application in analyzing the dynamics of metastable Markov open quantum systems~\cite{Brown2024}, we believe that {the SMP perspective} still has potential for investigating other nonequilibrium statistical issues. In this work, we report a new case: {calculating the LDs of the first passage time statistics (LDs-FPTS) in the same open quantum systems.}  

In contrast to the LDs-CS, which describe nonequilibrium large fluctuations in the counting variables at fixed large times, the LDs-FPTS describe large fluctuations in time as the same counting variables reach large threshold values~\cite{Budini2014,Roldan2015,Saito2016,Garrahan2017,Gingrich2017,Ptaszyifmmodenelsenfiski2018,Rudge2019}. Although the two types of statistics are usually  distinct~\cite{Ptaszyifmmodenelsenfiski2018,Rudge2019}, several studies have demonstrated that their LDs are indeed conjugated; that is, their respective scaled cumulant generating functions (SCGFs) are inverse functions of each other. This intriguing conclusion was obtained earlier by Budini {\it et al.}, who investigated the equivalence between trajectory ensembles in the classical master equations and QMEs~\cite{Budini2014}. In their paper, the time-extensive random variables are restricted to be nonnegative and strictly nondecrease with time, {\it e.g.}, the dynamical activity, which is the total number of jumps (collapse) in a classical (quantum) trajectory~\cite{Maes2008,Garrahan2010,Garrahan2017}. Later, Gingrich and Horowitz demonstrated that the conjugation also holds for the other type of time-extensive random variables, the currents in the classical mater equations~\cite{Gingrich2017}. 
Notably, the currents are asymmetric under time reversal, {\it e.g.}, the net difference in the number of transitions from one classical state to another versus the opposite. {In this work, we refer to these two types of random variables as simple counting variables and current-like variables, respectively. The latter may be positive or negative, and they do not need to be strictly asymmetric under time reversal. If not distinguished, we refer to all the time-extensive random variables as counting variables. The reason is that they are intimately related to counting.} A significant consequence of the conjugation between the LDs-FPTS and LDs-CS is the finding of another set of conjugations between the thermodynamic uncertainty relation (TUR) and the kinetic uncertainty relation (KUR) in the two types of statistics~\cite{Barato2015,Gingrich2016,Garrahan2017,Gingrich2017}. 
	  
The goal of this work is to demonstrate that the conjugation relationship between the LDs-FPTS and LDs-CS holds in the specific open quantum systems as well. The major motivation is the completeness of the theory. {Here, the counting variables are general: they are allowed to be positive or negative, rather than being restricted only to being nonnegative, and to increase or decrease with time~\cite{Budini2014,Garrahan2017}; they do not need to exhibit time-reversal asymmetry. Budini {\it et al.} developed a theory for the simple counting variables in open quantum systems. Hence, their theory is unsuitable for the current-like variables. In spirit, this work is more closely aligned with a generalization of Gingrich and Horowitz's work to the quantum regime. In addition, we present methods for directly calculating the LDs-FPTS. At first glance, such an effort seems unnecessary because the conjugation indicates that the SCGF of the FPTS (SCGF-FPTS) is the inverse function of the SCGF of the CS (SCGF-CS), whereas the latter can be calculated via the full counting statistics (FCS)~\cite{Mollow1975,Levitov1996,Lebowitz1999,Bagrets2003,Esposito2009,Garrahan2010,Landi2024} or the SMP method~\cite{Liu2022}. In the FCS, the calculation involves finding the largest real eigenvalues of the tilted generators of the QME. However, even in simple open quantum systems, this task is analytically intractable~\cite{Liu2023a}. Numerical schemes are often employed~\cite{Giardina2006,Lecomte2007,Bruderer2014,Cavallaro2016}. Notably, the conjugation does not imply that solving for the LDs-CS is necessary for obtaining the LDs-FPTS. Thus, we cannot exclude the possibility that directly solving for the latter is easier, and there may even be analytical solutions in some open quantum systems. }

The remainder of this paper is organized as follows. {In Sec.~(\ref{section2}), we review the SMP method for the CS in a specific class of open quantum systems. Several previous formulas are rederived to make them applicable to the FPTS. In particular, the region of convergence (ROC) for the joint Laplace transform and $z$-transform is carefully examined. Building on the foundation laid in the preceding section, in Sec.~(\ref{section3}), we demonstrate that the conjugation relationship between the LDs-FPTS and LDs-CS holds exactly in these open quantum systems. Moreover, for specific types of the counting variables, we present corresponding calculation methods for the SCGFs-FPTS. In Sec.~(\ref{section4}), a driven quantum two-level system (TLS) is used to illustrate the theoretical results, where we analytically solve for several SCGFs-FPTS and apply them to investigate quantum violations of the classical uncertainty relations.} Section~(\ref{section5}) concludes this work.

\section{Semi-Markov process method for counting statistics in open quantum systems}
\label{section2}
{We let $\rho(t)$ be the reduced density matrix of an open quantum system.} Under appropriate conditions, the dynamics of the system is described by the  QME~\cite{Davies1974,Lindblad1976,Gorini1976}
\begin{eqnarray}
	\label{MQME}
	\partial_t \rho(t)=-{\rm i}[H,\rho(t)]+\sum_{\alpha=1}^M r_\alpha\left( A_\alpha\rho(t)A^\dag_\alpha -\frac{1}{2}\left\{A^\dag_\alpha A_\alpha,\rho(t)\right\}\right) ,
\end{eqnarray}
where the Planck constant $\hbar$ is set to 1, $H$ denotes the Hamiltonian of the quantum system, $A_\alpha$ is the Lindblad operator or jump operator, and the nonnegative rates, $r_\alpha$,  for $\alpha=1,\cdots,M$, represent certain correlation characteristics of the environment surrounding the quantum system.

\subsection{Semi-Markov process in quantum jump trajectories}
\label{section2A}
\begin{figure}
	\includegraphics[width=1\columnwidth]{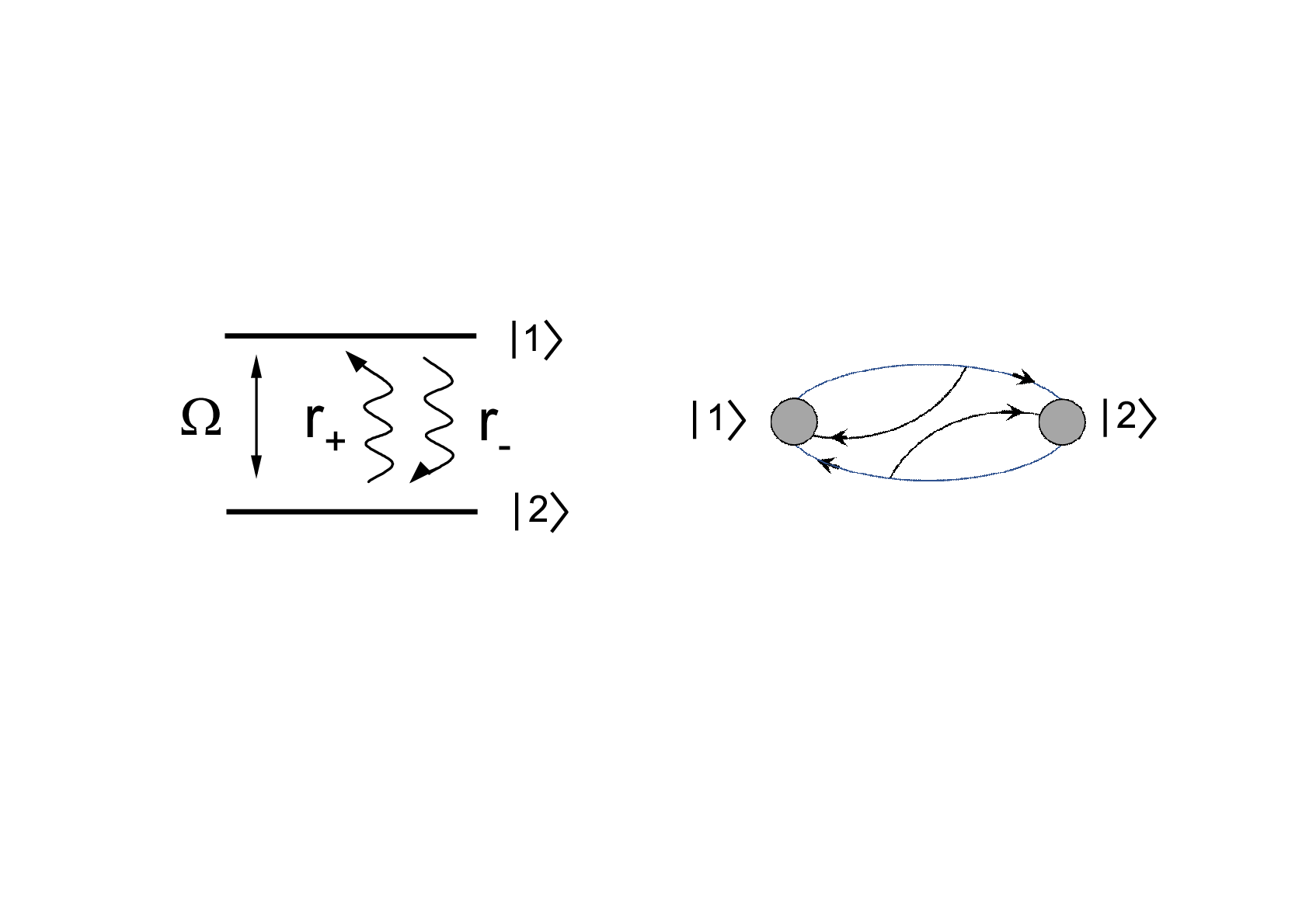}
	\caption{Left: A schematic representation of an externally driven quantum TLS surrounded by a finite-temperature environment. Right: A schematic representation of the QJTs of the TLS, in which a curve without an arrow denotes a deterministic piece of the wave function of the system and a curve with an arrow denotes a random collapse. From the SMP perspective, the starting and ending gray dots of the curves also denote the starting and collapsed quantum states of a WTD, respectively.}
	\label{fig1}
\end{figure}

The QME~(\ref{MQME}) can be unraveled into QJTs~\cite{Mollow1975,Srinivas1981,Zoller1987,Dalibard1992,Molmer93,Carmichael1993,Plenio1998,Breuer2002,Wiseman2010}. These trajectories, which describe the evolution of the wave functions of single quantum systems, consist of deterministic pieces and random wave function collapses. The former are the solutions to nonlinear Schr$\ddot{o}$dinger equations. The latter indicate that the systems collapse to certain quantum states. As in the case we were previously interested in, the set of collapsed states is assumed to be fixed and finite~\cite{Liu2022}. We let the elements of the set be $\phi_\alpha$, for $\alpha=1,\cdots, M$. Figure~\ref{fig1} shows this scenario in an externally driven TLS: the gray dots on the right represent the collapsed quantum states $|1\rangle$ and $|2\rangle$, the curves without arrows denote that the wave functions of a single system start from these states and continuously evolve, and the curves with arrows denote the random collapses of the wave functions to these states. If the entire evolution of the wave functions is traced, the QJT is essentially Markovian. However, if we focus on the collapsed states and use random time intervals $\tau$ to replace the deterministic pieces between successive collapses, the QJTs can be viewed as realizations of an SMP~\cite{Liu2022}. The components of the SMP include the waiting time densities (WTDs) and survival distributions (SDs)~\cite{Ross1995}, which are as follows:
\begin{eqnarray}
	\label{waitingtimedensity}
	p_{\beta|\alpha}(\tau)&=&r_\beta\parallel A_\beta e^{-{\rm i}\tau \tilde H}\phi_\alpha \parallel^2,\\ 
	S_\alpha(\tau)&=&\parallel e^{-{\rm i}\tau \tilde H}\phi_\alpha\parallel^2,
	\label{survivaldistribution}
\end{eqnarray}
respectively~\cite{Breuer2002}. Here, the non-Hermitian Hamiltonian is  
\begin{eqnarray}
	\tilde H=H-\frac{\rm i}{2}\sum_{\alpha=1}^M r_\alpha A_\alpha^\dag A_\alpha.
\end{eqnarray}
Equation~(\ref{waitingtimedensity}) is the probability density of the wave function starting from the collapsed state $\phi_\alpha$, continuously evolving, and collapsing in state $\phi_\beta$ until time $\tau$. Equation~(\ref{survivaldistribution}) is the probability of the wave function evolving successively until time $\tau$ without collapse.

\subsection{Counting variables} 
\label{section2B} 
Consider a QJT of a quantum system with $N$ collapses. The trajectory starts from a fixed quantum state $\phi_\eta$ and ends in the final collapsed state $\phi_\gamma$, where both $\eta$ and $\gamma$ are elements of the set $(1,\cdots,M)$. We represent the QJT as \begin{eqnarray}
\label{trajectorydefinition}
{\bf X}(t,N)=(\gamma,\alpha_{N-1},\cdots, {\alpha_2},{\alpha_1},\eta),
\end{eqnarray}
where $t$ is the total time of quantum evolution and $\phi_{\alpha_i}$ is the state that collapses at time $t_i$, for $i=0,1,\cdots, N$, with $\alpha_0=\eta$ and $\alpha_N=\gamma$. These states are arranged from left to right in chronological order. The time-extensive counting variable in the trajectory is defined as    
\begin{eqnarray}
	\label{countingquantity}
	C[{\bf X}(t,N)]=\sum_{i=N}^1 w_{\alpha_{i}\alpha_{i-1}}.  
\end{eqnarray}
Here, the coefficient $w_{\alpha_i\alpha_{i-1}}$ is a weight determined by the adjacent collapsed states $\phi_{\alpha_i}$ at times $t_i$ and $t_{i-1}$.  

Equation~(\ref{countingquantity}) is general enough to encompass the counting variables mentioned earlier. 
For example, if the weights $w_{\alpha\beta}$, for $\alpha,\beta=1,\cdots,M$, are set uniformly to $1$, the variable represents the dynamical activity~\cite{Maes2008,Garrahan2010}. If the weights are antisymmetric, that is, $w_{\alpha\beta}=-w_{\alpha\beta}$, the variable mimics the currents in the classical master equations. Finally, if the weights depend solely on $\phi_{\alpha_i}$, such that $w_{\alpha_i\alpha_{i-1}}\equiv w_{\alpha_i}$, the variable corresponds to that of interest in the FCS~\cite{Esposito2009,Landi2024}. The weights are arbitrary; however, for the purpose of the FPTS in this work, we concentrate on specific weights that are equal to $\pm 1$ or $0$. Consequently, the values of Eq.~(\ref{countingquantity}) are always discrete integers, and their changes over time are equal to $\pm 1$ or 0.

\subsection{Moment generating function of counting statistics}
\label{section2C}
An important property in the CS is the LDs of the counting variables~(\ref{countingquantity}) in the large time limit, $\it i.e.$, the SCGF-CS 
\begin{eqnarray}
	\label{SCGFdefinitioncountingstatistics}
	\varphi(\lambda)=\lim_{t\rightarrow\infty}\frac{1}{t}\ln Z_C(\lambda|t), 
\end{eqnarray}
where  
\begin{eqnarray}
	\label{MGFforcountingstatistics}
	Z_C(\lambda|t)=\langle e^{-\lambda C}\rangle
\end{eqnarray}
is the moment generating function of the CS (MGF-CS), $\lambda$ is a real number, and $\langle\cdots\rangle$ indicates the average over the distributions of the counting variables~\cite{Touchette2008}. To achieve a feasible calculation method for Eq.~(\ref{SCGFdefinitioncountingstatistics}), we first solve for the MGF-CS. 

To this end, we express the probability distribution of the QJT~(\ref{trajectorydefinition}) in terms of the WTDs~(\ref{waitingtimedensity}):     
\begin{eqnarray} 
\label{probdistributionQJT}
{\cal P}[{\bf X}(t,N)]= S_\gamma(t-t_N)p_{\gamma|\alpha_{N-1}}(t_N-t_{N-1}) \cdots p_{\alpha_2|\alpha_1}(t_2-t_1)p_{\alpha_1|\eta}(t_1). 
\end{eqnarray}
We let $T_{\gamma\eta}(n|t)$ be the joint transition probability that the quantum system starts from state $\phi_\eta$, the last collapsed state is $\phi_\gamma$, and the counting variable~(\ref{countingquantity}) equals $n$ at time $t$, simultaneously. On the basis of Eq.~(\ref{probdistributionQJT}), the joint probability is formally expressed as 
\begin{eqnarray}
\label{jointtransitionprob}
T_{\gamma\eta}(n|t)=\sum_{N=0}^\infty\sum_{{\bf  X}(t,N)} {\cal P}[{\bf X}(t,N)] \delta_{nC[{\bf  X}(t,N)]}.
\end{eqnarray} 
Here, the second summation is shorthand notation: it represents the sums over all possible intermediate collapsed states $\phi_{\alpha_i}$ and integrals over all possible times $t_i$, for $t\ge t_N\ge \cdots\ge t_1\ge 0$. Equation~(\ref{jointtransitionprob}) is complex but can be significantly simplified by taking the Laplace transform of time $t$ and the $z$-transform of the count $n$ consecutively. We obtain   
\begin{eqnarray}
\label{transformedjoinedtransitionprobs}
\bar {\hat T}_{\gamma\eta}(z|\nu)&=&\int_0^\infty \sum_{n=-\infty}^{+\infty}T_{\gamma\eta}(n|t)z^{-n}e^{-\nu t}dt\nonumber\\
&=&{\hat S}_\gamma(\nu) \sum_{N=0}^\infty \sum_{\alpha_{N-1}=1}^M\cdots\sum_{\alpha_{1}=1}^M {\hat p}_{\gamma|\alpha_{N-1}}(\nu)z^{-w_{\gamma\alpha_{N-1}}}\cdots  {\hat p}_{\alpha_2|\alpha_1}(\nu)z^{-w_{\alpha_2\alpha_1}} {\hat p}_{\alpha_1|\eta}(\nu)z^{-w_{\alpha_1\eta}}.\nonumber\\
&=&	[\hat{{\mathbb S}}]_{\gamma\gamma}\sum_{N=0}^\infty \sum_{\alpha_{N-1}=1}^M\cdots\sum_{\alpha_{1}=1}^M	[{\mathbb W}]_{\gamma\alpha_{N-1}}\cdots	[{\mathbb W}]_{\alpha_2\alpha_1}[{\mathbb W}]_{\alpha_1\eta}\nonumber \\
&=&\left[\hat{{\mathbb S}}(\nu)\sum_{N=0}^\infty {\mathbb W}^N(\nu,z)\right]_{\gamma\eta}
=\left[ \hat{{\mathbb S}}(\nu)\hspace{0.1cm}
\frac{1}{\mathbb{ I - W}(\nu,z)}\right]_{\gamma\eta}.
\end{eqnarray}
Throughout the rest of this work, we use a ``hat" over a symbol to denote the Laplace transform with respect to time $t$, and use a ``bar" over a symbol to denote the z-transform of counting $n$. In the third line of  Eq.~(\ref{transformedjoinedtransitionprobs}), we define three $M\times M$ matrices: the unit matrix ${\mathbb I}$, and matrices $\mathbb W$ and $\mathbb S$ with following elements: 
\begin{eqnarray}
	\label{Wmatrix}
	[{\mathbb W}]_{\alpha\beta}=\hat p_{\alpha|\beta}(\nu)z^{- w_{\alpha\beta}}  
\end{eqnarray} and \begin{eqnarray}
	\label{Survivalmatrix}
	[\hat{{\mathbb S}}]_{\alpha\beta}={\hat S}_\alpha(\nu)\delta_{\alpha\beta},
\end{eqnarray}
respectively, where $\delta_{\alpha\beta}$ is the Kronecker delta, for $\alpha,\beta=1,\cdots,M$. The last equation is a result of applying the Neumann series, and to ensure the convergence of the series, the spectral radius of $\mathbb W$ must be less than $1$~\cite{Meyer2023}. This restriction, combined with convergence conditions for the Laplace transforms of WTDs, specifies the ranges of $\nu$ and $z$ for which these transformed joint transition probabilities converge. Here, the values of $\nu$ and $z$ can be complex numbers. Then, using Eq.~(\ref{transformedjoinedtransitionprobs}), and assuming that the initial probability of the quantum system starting from state $\phi_\eta$ is $P_\eta(0)$, we derive the Laplace transform of the MGF-CS:
\begin{eqnarray}
	\label{LaplacetransformedMGF}
	\hat Z_C(\lambda|\nu)&=&\sum_{\gamma=1}^M\sum_{\eta=1}^M \bar {\hat T}_{\gamma\eta}(z|\nu) P_\eta(0)\nonumber \\
	&=& {\bf 1}^{\text{T}} \hat{{\mathbb S}}(\nu)\hspace{0.1cm}
	\frac{1}{\mathbb{ I - W}(\nu,z)} {\bf P}(0).
\end{eqnarray}  
Here, $z=\exp(\lambda)$, the superscript ``${\text T}$" denotes the transpose, and we use bold font to denote an $M\times 1$ column matrix, {\it e.g.}, $[{\bf P}]_\eta=P_\eta$, and ${\bf 1}=(1\cdots,1)$. It should be noted that the value of $z$ in Eq.~(\ref{LaplacetransformedMGF}) differs from that in  Eq.~(\ref{transformedjoinedtransitionprobs}); the former is real and positive.   
 
\subsection{Region of convergence for joint transform}
\label{section2D}
Before moving on to the LDs-CS, we first carefully examine the convergence of the joint transform,  Eq.(\ref{transformedjoinedtransitionprobs}). To begin with, we imagine a high-dimensional space defined by two complex variables, $\nu$ and $z$. Within this space, we hypothesize that points characterized by spectral radii less than 1 are densely distributed across a particular region. Because LD theory is concerned primarily with MGFs with real parameters, such as Eq. (\ref{LaplacetransformedMGF}), we consider a cross-section. This cross-section is defined by the intersection of a two-dimensional real $\nu$-$z$ plane with the aforementioned high-dimensional region, provided that this intersection is nonempty. All the points within the intersection are evidently capable of ensuring the convergence of Eq. (\ref{transformedjoinedtransitionprobs}). We refer to this cross-section as the ROC.

Next, we demonstrate that the boundary of the ROC consists of points that satisfy a rational equation involving two real variables $\nu$ and $z$:   
\begin{eqnarray}
	\label{equationofpoles}
	\det[{\mathbb I}-{\mathbb W}(\nu,z)]=0.
\end{eqnarray}
Note that this result relies on the assumption that the weights in Eq.~(\ref{countingquantity}) are equal to $\pm 1$ or $0$. Since the proof of this equation involves properties of matrix eigenvalues, we have detailed the procedure in Appendix A. Although Eq.~(\ref{equationofpoles}) defines the boundary equation of the ROC, throughout the rest of this work, we refer to it as the equation of poles. The reason is as follows: for a given real value of $z$, the roots of the equation, $\nu_k(z)$, for $k=1,\cdots$, represent the poles of the Laplace transform~(\ref{transformedjoinedtransitionprobs}) when $z$ is held constant. Similarly, for a given real value of $\nu$, the roots of the equation, $z_m(\nu)$, for $m=1,\cdots$, represent  the poles of the  $z$-transform~(\ref{transformedjoinedtransitionprobs}) when $\nu$ is held constant. Naturally, the points on the boundary are part of the set of real solutions to Eq.~(\ref{equationofpoles}).

Then, Eq.~(\ref{transformedjoinedtransitionprobs}) is rational with respect to the  variables $\nu$ and $z$. Consequently, either its inverse Laplace transform or inverse $z$-transform is a linear combination of exponentials~\cite{Oppenheim1997}. This characteristic imposes two constraints on the shape of the ROC. First, in the real $\nu$-$z$ plane, the maximum number of intersections between a horizontal line at a given $z$ value or a vertical line at a given $\nu$ value and the boundary of the ROC must not exceed two. To illustrate, we consider a horizontal line at a fixed $z$ value. If there were more than two intersections, it would imply the existence of at least two separate intervals of $\nu$ for which the Laplace transform in Eq. (\ref{transformedjoinedtransitionprobs}) converges. This contradicts the established property that the convergence range for the Laplace transform of a sum of exponentials is bounded by exactly two poles, as detailed in property 7, Section 9.2 of Ref.~\cite{Oppenheim1997}. The same logic applies to the case of a vertical line at a fixed $\nu$ value. Second, points on the boundary of the ROC are distinctive. We consider a horizontal line intersecting the ROC in the $\nu$ direction. We specifically refer to the portion of the line within the ROC as $\text{ROC}(z)$, which denotes the range of $\nu$ values for which the Laplace transform, Eq.~(\ref{transformedjoinedtransitionprobs}), converges when $z$ is held constant. We assert that the real part of any root $\nu_k(z)$, excluding those at the intersections with the boundary, $\text{Re}[\nu_k(z)]$, must lie outside of $\text{ROC}(z)$ (\textit{first statement}). This is because the convergence range of a Laplace transform is determined solely by the real part of its poles. Therefore, any pole or root violating this statement would result in a reduction of the original ROC in the $\nu$ direction. Similarly, we consider a vertical line intersecting the ROC in the $z$ direction. We refer to the portion of the line within the ROC as $\text{ROC}(\nu)$, which denotes the range of $z$ values for which the $z$-transform,  Eq.~(\ref{transformedjoinedtransitionprobs}), converges when $\nu$ is held constant. We assert that the modulus of any root $z_m(\nu)$, excluding those at the intersections with the boundary, $|z_m(\nu)|$, must be either smaller or larger than the magnitudes of the points within $\text{ROC}(\nu)$ (\textit{second statement}). This is because the convergence range of a $z$-transform depends solely on the modulus of its poles. Hence, any pole or root violating this statement would also lead to a reduction of the original ROC in the $z$ direction.
     
Finally, the joint transition probabilities given in Eq.~(\ref{jointtransitionprob}) are causal, meaning that they are zero for $t<0$. Since Eq.~(\ref{transformedjoinedtransitionprobs}) is rational in $\nu$, the convergence range of $\nu$ is to the right of its rightmost pole~\cite{Oppenheim1997}. This implies that if we consider a horizontal line intersecting with the ROC in the $\nu$ direction while $z$ is held constant, there will be only one  intersection point. Consequently, the $\text{ROC}(z)$ extends from this intersection point to infinity in the $\nu$ direction. This conclusion further reinforces the first constraint on the shape of the ROC.  

\begin{figure}
	\includegraphics[width=1\columnwidth]{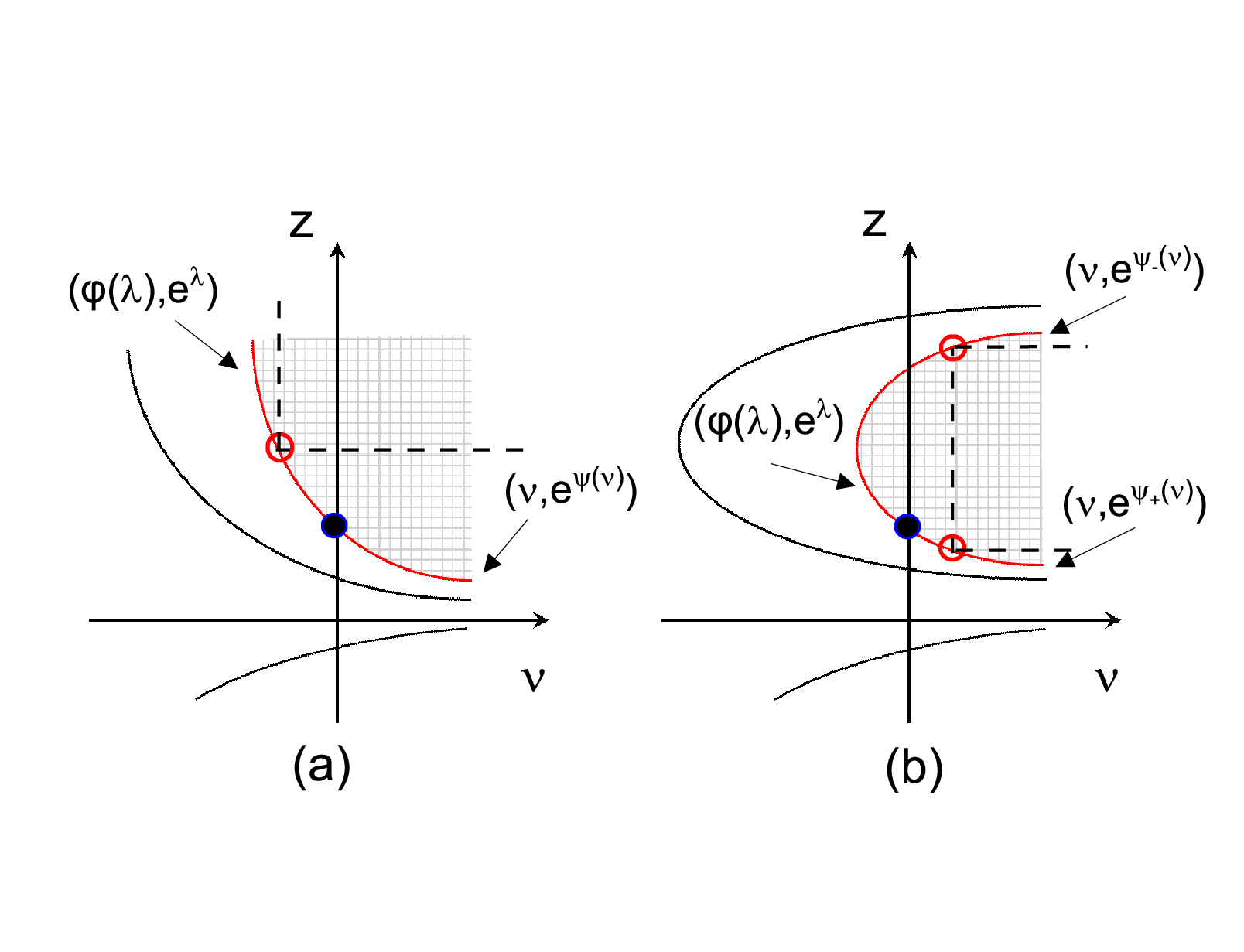}
	\caption{ The shaded areas in (a) and (b) are schematic representations of the ROC for Eq.~(\ref{transformedjoinedtransitionprobs}). The curves illustrate the real solutions to the equation of poles~(\ref{equationofpoles}). The two rightmost red curves signify the boundaries of the ROC. Notably, both curves pass through point $(0,1)$, which is denoted by the blue solid dot. This is because the rank of the matrix ${\mathbb I}-{\mathbb W}$ given in Eq.~(\ref{equationofpoles}) at this point is less than $M$. The dashed horizontal and vertical line segments within the shaded areas represent $\text{ROC}(z)$ and $\text{ROC}(\nu)$, respectively. The difference between panels (a) and (b) is in the extent of $\text{ROC}(\nu)$. In panel (a), $\text{ROC}(\nu)$ extends from the boundary's $z$ value to infinity in the $z$-direction. In contrast, in panel (b), $\text{ROC}(\nu)$ is bounded by the two $z$ values on the boundary for a given $\nu$ value, as shown by the red open circles. It should be noted that the range of $z$ is $(0,+\infty)$, whereas this is not the case for $\nu$. 
}
	\label{fig2}
\end{figure}

When we synthesize all the previously discussed knowledge about the ROC for the joint transform, we can schematically represent it in Fig.~\ref{fig2}; see the shaded areas. These areas extend to infinity in the $\nu$ direction. Furthermore, we categorize them into two classes on the basis of their characteristics in the $z$ direction: one class extends to infinity, whereas the other class is bounded by two real poles on the boundary.

We conclude this section by examining the ROC for the Laplace transform of the MGF-CS, Eq. (\ref{LaplacetransformedMGF}). Since this equation is directly derived from Eq. (\ref{transformedjoinedtransitionprobs}), its ROC is identical to that of the latter equation. Specifically, because the LD principle, Eq. (\ref{SCGFdefinitioncountingstatistics}), applies to all values of $z=\exp(\lambda)$, which are greater than zero, this ROC must be situated in the upper half of the $\nu$-$z$ plane. Consequently, the shaded areas in Fig.~\ref{fig2} also represent the ROC for Eq. (\ref{LaplacetransformedMGF}). Conversely, this explains why the areas representing the ROC for Eq.~(\ref{transformedjoinedtransitionprobs}) are positioned in the upper half of the plane. 

\subsection{Scaled cumulant generating function of counting statistics}
\label{section2E}
As previously mentioned, the Laplace transform of the MGF-CS, Eq.~(\ref{LaplacetransformedMGF}), is rational in $\nu$. By performing a partial-fraction expansion, we express it as follows~\cite{Oppenheim1997}: 
\begin{eqnarray}
	\label{Laplacetransformgeneratingfunctionexpanded}
	{\hat Z}_C(\lambda|\nu) \sim \sum_{k=1} \frac{1}{\nu-\nu_k(z)},
\end{eqnarray}
where $\nu_k$ represent the distinct roots or poles of Eq.~(\ref{equationofpoles}) in $\nu$, which can be complex numbers. For simplicity in our mathematical representation, we do not explicitly list the coefficients in front of each term in Eq. (\ref{Laplacetransformgeneratingfunctionexpanded}), nor do we consider the possibility of multiple roots. To uniquely determine the inverse Laplace transform of Eq.~(\ref{Laplacetransformgeneratingfunctionexpanded}), which is the MGF-CS  $Z_C(\lambda|t)$, we must specify the range of $\nu$ for the convergence of the Laplace transform $\hat Z_C(\lambda|\nu)$ when $z$ is held constant, denoted as  $\text{ROC}(z)$. As depicted in Fig.~\ref{fig2}, this range extends from the boundary's $z$ value to infinity in the $\nu$ direction, represented by the horizontal line segments in the figure. According to the first statement that the real parts of all the roots $\nu_k(z)$ lie outside of $\text{ROC}(z)$, the inverse Laplace transform is a linear combination of exponentials: 
\begin{eqnarray}
	\label{asymptoticMGFforcountingstatistics}
	Z_C(\lambda|t)\sim \sum_{k=1} e^{t\nu_k (z)}.
\end{eqnarray}
Following the LD principle, Eq.~(\ref{SCGFdefinitioncountingstatistics}), we derive the SCGF-CS~(\ref{countingquantity}):  
\begin{eqnarray}
	\label{SCGFformulacountingstatistics}
	\varphi(\lambda)=\max_k\{\text{Re}[\nu_k(z)]\}.
\end{eqnarray}
Due to the first statement, this SCGF is equivalent to the value of $\nu$ at the boundary point for a given $z$ value. Conversely, a point $(\nu,z)$ on the boundary can be represented by the SCGF-CS as $(\varphi(\lambda),\exp(\lambda))$. 

Two comments are in order. First, Eqs.~(\ref{transformedjoinedtransitionprobs})-(\ref{LaplacetransformedMGF}), (\ref{equationofpoles}), and (\ref{SCGFformulacountingstatistics}) are attributed to Andrieux and Gaspard, who originally derived these formulas in their study of the current fluctuation theorem within classical SMPs~\cite{Andrieux2008}. We have extended these formulas to apply to the specific open quantum systems. It is important to note that the discussion on the ROC for the joint transform is being presented for the first time in this work. Second, in a previous paper, we introduced a matrix $\mathbb G$ for calculating the Laplace transform of the MGF-CS for the counting variables~\cite{Liu2022}. The relationship between the matrices $\mathbb G$ and $\mathbb{W}$ is detailed in Appendix B.

\section{Large deviations of first passage time statistics}
\label{section3}
\subsection{Moment generating function of first passage time statistics}
The FPTS focuses on the probability distribution $F({\cal T}|n)$, which describes the possibility that the counting variable~(\ref{countingquantity}) first passes a threshold value $n$ at time ${\cal T}$~\cite{Budini2014,Roldan2015,Saito2016,Garrahan2017,Gingrich2017,Ptaszyifmmodenelsenfiski2018,Rudge2019}. We highlight the distinction between ${\cal T}$ and the previously mentioned $t$: the former is a random variable, while the latter represents the total time of the quantum evolution and is fixed. Similar to the CS, we first solve for the MGF-FPTS as follows:  
\begin{eqnarray}
	\label{MGFforFPTstatistics} Z_F(\nu|n)=\left\langle e^{-\nu {\cal T}}\right\rangle
\equiv {\hat F}(\nu|n), 
\end{eqnarray}
where $\nu$ is a real number. The final expression of Eq.~(\ref{MGFforFPTstatistics}) follows the convention for the Laplace transform of time. We define the distribution  $F_{\alpha}({\cal T}|n)$, which represents the joint FPT distribution of the counting variable~(\ref{countingquantity}) passing through a threshold value $n$ at time ${\cal T}$ and the last collapsed state of the system being $\phi_\alpha$. It is clear that 
\begin{eqnarray} 
F({\cal T}|n)=\sum_{\alpha=1}^M F_\alpha({\cal T}|n).
\end{eqnarray} 
Saito and Dhar, and Ptaszynski have derived a probability relationship between the FPT distributions and the joint  transition probabilities, Eq.~(\ref{jointtransitionprob})~\cite{Saito2016,Ptaszy2018}. In the complex frequency domain, this relationship takes a concise form:   
\begin{eqnarray}
	\label{SatioDharequation}
	\sum_{\eta=1}^{M}	{\hat T}_{\gamma \eta}(n|\nu)P_{\eta}(0) =\sum_{\eta=1}^{M} \hat{T}_{\gamma\eta}(0|\nu){\hat F}_\eta(\nu|n). 
\end{eqnarray}
Importantly, although this equation was originally derived for Markov processes~\cite{Saito2016,Ptaszy2018}, it is also valid for SMPs. This validity stems from the fact that, in the specific class of open quantum systems considered in this work, a wave function collapse implies that the subsequent quantum process is independent of the system's history prior to collapse. However, in general open quantum systems, the systems retain their history even if quantum collapses occur~\cite{Landi2024}. Under such circumstances, Eq. (\ref{SatioDharequation}) may not be applicable.

We apply the $z$-transform of $n$ to both sides of Eq.~(\ref{SatioDharequation}). Furthermore, we define an $M\times M$ matrix $\bar {\hat{\mathbb T}}$ and an $M\times 1$ column matrix $\bar {\hat{\bf F}}$, with elements $[{\mathbb T}]_{\gamma\eta}=\bar {\hat T}_{\gamma\eta}$ and $[\bar {\hat {\bf F}}]_\eta $$=\bar{\hat{F}}_\eta$, respectively. This $z$-transform results in a matrix equation:  
\begin{eqnarray}
	\label{SatioDharequationtransformedmatrixform}
	\bar{\hat{{\mathbb T} }}(z|\nu){\bf P}(0)={\hat{{\mathbb T} }}(0|\nu){\bar {\hat{\bf F}}}(\nu|z),
\end{eqnarray}  
Finally, by substituting Eq.~(\ref{transformedjoinedtransitionprobs}) into Eq.~(\ref{SatioDharequationtransformedmatrixform}), we obtain the $z$-transform of the MGF-FPTS,     
\begin{eqnarray}
	\label{ztransformedMGF}
	{\bar Z}_{F}(\nu|z)&=&{\bf 1}^{\text{T}}{\bar {\hat{\bf F}}}(\nu|z)\nonumber\\
	&=&{\bf 1}^{\text T}{\hat{{\mathbb T}}^{-1}(0|\nu)}\hat{{\mathbb S}}(\nu) \frac{1}{ {\mathbb I}-{\mathbb W}(\nu,z) }{\bf P}(0).
\end{eqnarray}
The derivation clearly shows that Eq.~(\ref{ztransformedMGF}) also stems from the joint transform,  Eq.~(\ref{transformedjoinedtransitionprobs}). Consequently, its ROC is exactly the same as that for the Laplace transform of the MGF-CS, Eq.~(\ref{LaplacetransformedMGF}).

\subsection{Scaled cumulant generating function of first passage time statistics }	
We are now ready to investigate the LDs-FPTS. The following discussion is analogous to that in Sec.~(\ref{section2D}). First, the structure of Eq.~(\ref{ztransformedMGF}) is similar to that of Eq.~(\ref{LaplacetransformedMGF}): the former is rational but in $z$. Consequently, by performing a  partial-fraction expansion, we can express Eq.~(\ref{ztransformedMGF}) as follows~\cite{Oppenheim1997}: 
\begin{eqnarray}
	\label{ztransformexpansion}
	{\bar Z}_F(\nu|z) \sim \sum_{m=1} \frac{1}{1-z_m(\nu)z^{-1}},
\end{eqnarray}
where $z_m(\nu)$ represent the distinct roots or poles of Eq.~(\ref{equationofpoles}) but in terms of $z$, which can be complex numbers. As with the CS, we do not explicitly list the coefficients in front of each term, nor do we consider the more general case with multiple roots. Next, we use Eq.~(\ref{ztransformexpansion}) associated with its $\text{ROC}(\nu)$ to obtain the desired inverse $z$-transform or the MGF-FPTS $Z_F(\nu|n)$. As shown in Fig.~\ref{fig2}, there are two classes of ROCs depending on their extents in the $z$-direction. These need to be discussed separately.

\subsubsection{Simple counting variables}
In the first scenario, as depicted in  Fig.~\ref{fig2}(a), for a given value of $\nu$, $\text{ROC}(\nu)$ extends from the boundary's $z$ value to infinity in the $z$ direction, that is, the vertical line segment shown in the panel. According to the second statement, the moduli of all the roots $z_m(\nu)$ must be equal to or smaller than the magnitudes of the points within $\text{ROC}(\nu)$. Hence, the inverse $z$-transform of Eq.~(\ref{ztransformexpansion}) is a linear combination of the right-sided exponentials: 
\begin{eqnarray}
\label{asymptoticMGFforFPTstatisticstype1}
	Z_F(\nu|n)\sim \sum_{m=1} (z_m(\nu))^n u(n),
\end{eqnarray}
where $u(n)$ is the unit step function. Under these conditions, the counting variables are simple: they are always positive and increase monotonically with time, such as the dynamical  activity~\cite{Budini2014,Garrahan2017}. Following the LD principle, this suggests that in the large $n$ limit, the SCGF-FPTS is given by 
\begin{eqnarray}
	\label{SCGFdefinitionFPTtype1}
	\psi(\nu)=
	\lim_{n\rightarrow +\infty} \frac{1}{n}\ln Z_F(\nu|n)=\ln \max_m\{ |z_m(\nu)| \}.  
\end{eqnarray}
Due to the second statement, this SCGF is equivalent to the value of $\ln z$ at the boundary point for the given $\nu$ value. Conversely, a point $(\nu,z)$ on the boundary can be represented by the SCGF-FPTS as $(\nu, ,\exp(\psi(\nu)))$. As clearly shown in Fig.~\ref{fig2}(a), the boundary point here is identical to that discussed in Sec.~(\ref{section2E}) when we examined the LDs-CS. This indicates that $\psi=\varphi^{-1}$. Therefore, we conclude that for the simple counting variables, the SCGFs for each FPTS and CS are inverse functions of each other in the specific open quantum systems~\cite{Budini2014,Garrahan2017}. 
 
\subsubsection{Current-like variables}
In the second scenario, as depicted in Fig.~\ref{fig2}(b), for a given value of $\nu$, $\text{ROC}(\nu)$ is bounded by two $z$ values on the boundary for that $\nu$ value. We denote these values as 
$z_+(\nu)$ and $z_-(\nu)$, with the latter being greater. According to the second statement, the modulus of any root  $z_m(\nu)$ must be either equal to or smaller than $z_+(\nu)$, or equal to or greater than $z_-(\nu)$. Then, on the basis of the established property of the $z$-transform ({\it e.g.}, property 4, Sec.10. 4 in Ref~\cite{Oppenheim1997}), the inverse $z$-transform of Eq.~(\ref{ztransformedMGF}) is  
\begin{eqnarray}
	\label{asymptoticMGFforFPTstatisticstype2}
	Z_F(\nu|n)\sim -\sum_m (z_m^-(\nu))^n u(-n-1)+ \sum_m (z_m^+(\nu))^n u(n).
\end{eqnarray}
Here, the roots are divided into two sets: $z_m^+(\nu)$ are the roots whose moduli are equal to or smaller than $z_+(\nu)$, while $z_m^-(\nu)$ are the roots whose moduli are equal to or greater than $z_-(\nu)$. Under these conditions, the counting variables~(\ref{countingquantity}) are current-like: they can be positive or negative and may increase or decrease with time, such as the entropy production~\cite{Horowitz2019}. Following the LD principle, Eq.~(\ref{asymptoticMGFforFPTstatisticstype2}) yields two SCGFs-FPTS: 
\begin{eqnarray}
	\label{SCGFdefinitionFPTtype2positive}
	\psi_+(\nu)&=&\lim_{n\rightarrow +\infty} \frac{1}{n}\ln Z_F(\nu|n)=\ln \max_m\{ |z_m^+(\nu)| \},
\end{eqnarray}
and
\begin{eqnarray}
	\label{SCGFdefinitionFPTtype2negative}
	\psi_-(\nu)&=&\lim_{n\rightarrow -\infty} \frac{1}{n}\ln Z_F(\nu|n)=\ln \min_m\{ |z_m^-(\nu)| \},
\end{eqnarray}
respectively. Due to the second statement, these two functions correspond to the values of $\ln z$ at the two boundary points for the given $\nu$ value. Conversely, the two points, $(\nu,z_+(\nu))$ and $(\nu,z_-(\nu))$ on the boundary, can be represented by the two SCGFs-FPTS as $(\nu, \exp(\psi_+(\nu)))$ and $(\nu, ,\exp(\psi_-(\nu)))$, respectively. As clearly shown in Fig.~\ref{fig2}(b), these two points are identical to the boundary points discussed in Sec.~(\ref{section2E}). Therefore, we obtain the desired result that $\psi_\pm=\varphi^{-1}$ for the current-like variables in the specific open quantum systems. It should be noted that here, the inverse function $\varphi^{-1}$ is defined with respect to the upper and lower curves, which are represented schematically in Fig.~\ref{fig2}(b).

We summarize several calculation methods for the SCGFs-FPTS. For the simple counting variables, we solve for all the roots of Eq.~(\ref{equationofpoles}) in $z$ for a given value of $\nu$. Then, according to Eq.~(\ref{SCGFdefinitionFPTtype1}), the functions are equal to the largest modulus of the roots. This method is analogous to that for the SCGFs-CS, where we solve for all the roots of Eq.~(\ref{equationofpoles}) but in $\nu$ for a given value of $z$, then, according to Eq.~(\ref{SCGFformulacountingstatistics}), these functions are equal to the largest real part of the roots. For the current-like variables, the situation is distinct. We still solve for all the roots of Eq.~(\ref{equationofpoles}) in $z$ for a given value of $\nu$. Then, according to 
Eqs.~(\ref{SCGFdefinitionFPTtype2positive}) and (\ref{SCGFdefinitionFPTtype2negative}), the SCGFs-FPTS are equal to the largest and smallest moduli of the two sets of roots. To determine these sets, we would need to specify the ROC for the joint transform in advance. However, if this is the case, this calculation method would be redundant, since the aforementioned discussion clearly indicates that the ROC already contains all the knowledge about the SCGFs. Therefore, this calculation method for the current-like variables is not meaningful. In this situation, there are two possible solutions. One is that if Eq.~(\ref{equationofpoles}) can be reduced to a quadratic equation in $z$, the difficulty of nonuniqueness of the ROC will not arise at all. The reason is that the region would be uniquely enclosed by the two real roots of the equation, as schematically represented in Fig.~\ref{fig2}(b). At first glance, this may seem too specific to have general applications. However, we demonstrate that this case indeed occurs in the typical TLSs. The other solution is apply the established conclusion that the SCGFs-FPTS and SCGFs-CS are inverse functions of each other. Because solving for the latter functions does not rely on knowledge of the ROC in advance, it is always feasible to obtain the former functions by performing an inverse function transformation. We refer to this method as the inverse function method. It is important to note that the inverse functions are bidirectional, meaning that we can solve for the SCGFs-CS from the SCGFs-FPTS if the latter are more accessible.  
   
\section{Quantum Two-Level System}
\label{section4}
Here, we use a TLS to illustrate the previous theoretical results and their applications. This system is driven by a resonant field and is surrounded by an environment with an inverse temperature $\beta$; see the schematic representation in Fig.~\ref{fig1}. In the interaction picture, the QME of the system is 
\begin{eqnarray}
	\label{QMEtwolevel}
	\partial_t\rho(t)&=&-{\rm i}\left[ H,\rho(t)\right ] + r_-[\sigma_-\rho(t)\sigma_+ -\frac{1}{2}\{\sigma_+\sigma_-, \rho(t) \}   ]\nonumber \\
	&&+r_+[\sigma_+\rho(t)\sigma_- -\frac{1}{2}\{\sigma_-\sigma_+, \rho(t) \}   ].
\end{eqnarray}
Here, $H=-\Omega(\sigma_- +\sigma_+)/2$ represents the interaction Hamiltonian between the TLS and the field, $\sigma_\pm$ represent the raising and lowering Pauli operators, and $\Omega$ represents the Rabi frequency. The coefficients $r_-=r_0(\bar{n}+1)$ and $r_+=r_0{\bar n}$ are the pumping and damping rates, respectively, where $r_0$ is the spontaneous decay rate and $\bar n$ is the Bose--Einstein distribution of the environment. These two rates satisfy the detailed balance condition $r_-=r_+\exp{(\beta\omega_0 )}$, where $\omega_0$ is the energy level difference. The excited state $|1\rangle$ and the ground state $|2\rangle$ are clearly the two collapsed states. For the system, the equation of poles~(\ref{equationofpoles}) is simply  
\begin{eqnarray}
	\label{equationofpolesforTLSgeneralweights}
	[1-\hat p_{1|1}(\nu)z^{-w_{11}}][1-\hat p_{2|2}(\nu)z^{-w_{22}}]- \hat p_{2|1}(\nu)\hat p_{1|2}(\nu)z^{-w_{12}-w_{21}}=0. 
\end{eqnarray}
The Laplace transforms of the WTDs for the TLS are exactly solved and shown at the end of Appendix C.

\subsection{Several exact large deviations for first passage time statistics}
\label{section4A}
We consider three counting variables and solve for their respective SCGFs-FPTS. The first is the number of collapses to the ground state $|2\rangle$, that is, the weights are $w_{21}=w_{22}=1$ and $w_{11}=w_{12}=0$. In quantum optics, collapse to the ground state indicates the emissions of photons. Therefore, this variable is also interpreted as a count of single photons~\cite{Plenio1998}. Since this is a simple counting variable with nonnegative values, its LD-CS is solved for by the FCS ~\cite{Budini2014,Garrahan2010}. 
By substituting the expressions $\hat p_{\alpha|\beta}$, for $\alpha,\beta=1,2$, into Eq.~(\ref{equationofpolesforTLSgeneralweights}), simple algebraic operations lead to a polynomial in $z$ for a given $\nu$ value:  
\begin{eqnarray}
\label{equationofpolesforTLScountingground}
[2r_+r_-\xi(\nu)+r_- \Omega^2]z^{-1}-2\xi(\nu)[\xi^2(\nu)+4\mu^2]+r_+\Omega^2=0, 
\end{eqnarray}
where $\xi(\nu)=\nu+r/2$, $r=r_-+r_+$, $16\mu^2=4\Omega^2-\delta^2$, and $\delta=r_--r_+$. Equation~(\ref{equationofpolesforTLScountingground}) has a single root. According to Eq.~(\ref{SCGFdefinitionFPTtype1}), the SCGF-FPTS is simply     
\begin{eqnarray}
	\label{SCGFofTLScountingground}	
	\psi(\nu)=\ln z(\nu)=\ln  \frac{2r_+r_-\xi(\nu)+r_- \Omega^2}{2\xi(\nu)[\xi^2(\nu)+4\mu^2]-r_+\Omega^2}. 
\end{eqnarray} 
If we choose a zero-temperature environment, $\it i.e.$, $\bar n=0$, and consider the specific parameters where $2\Omega=r_-$, Eq.~(\ref{SCGFofTLScountingground}) is reduced to 
$\psi(\nu)=-3\ln(1+\nu/\Omega)$, which is the result in Ref.~\cite{Budini2014}. 

{
The second counting variable is simpler than the first: it counts the number of collapses where the TLS starts from the excited state $|1\rangle$ and ends in the ground state $|2\rangle$. The corresponding weights are $w_{21}=1$, and $w_{22}=w_{11}=w_{12}=0$. This variable mimics that of the classical TLS, where classical jumps from a higher to a lower energy level are counted~\cite{Garrahan2017}. Because this is still a simple counting variable, by repeating the previous process, we derive another polynomial in $z$:  
\begin{eqnarray}
\label{equationofpolesforTLScounting12only}
	\left[\xi(\nu)(\xi^2(\nu)+4\mu^2)-\frac{1}{4}r\Omega^2\right]^2-\frac{1}{16}\delta^2\Omega^4-r_+r_-\left[\left(\xi^2(\nu)+\frac{1}{2}\Omega^2\right)^2-\frac{1}{4}\delta^2\xi^2(\nu)\right]z^{-1}=0,
\end{eqnarray}
and the SCGF-FPTS is given by  
\begin{eqnarray}
	\label{SCGFofTLScounting12only}	
	\psi(\nu)=\ln\frac{r_+r_-\left[(\xi^2(\nu)+\Omega^2/2)^2-\delta^2\xi^2(\nu)/4\right]}{\left[\xi(\nu)(\xi^2(\nu)+4\mu^2)-\Omega^2 r/4\right]^2-\Omega^4\delta^2/16}.
\end{eqnarray} 
If we set $\Omega$ to zero, the seemingly complicated formula  simplifies significantly to the SCGF-FPTS for the incoherent classical TLS: $\psi(\nu)=\ln\left[ r_+r_-/(\nu+r_+)(\nu+r_-)\right]$~\cite{Garrahan2017}. In fact,  Eq.~(\ref{SCGFofTLScountingground}) also reduces to the same result. The physical reason is that, in the absence of driving, quantum coherence rapidly dissipates, causing the quantum TLS to degrade into its classical counterpart in the large time limit. We emphasize that, although the second counting variable is very simple, the FCS based on the tilted QME is incapable of calculating the SCGF-CS~\cite{Liu2022}. This limitation is one of the motivations that drove us to develop the SMP method for the specific open quantum systems. 
  
Finally, we consider a current-like variable. We assign the weights as $w_{21}=w_{22}=1$ and $w_{11}=w_{12}=-1$.} Now, the values of the counting variable can be positive or negative. If the variable is multiplied by $\omega_0$, the resulting quantity represents heat production in a QJT~\cite{Breuer2003,Liu2016a}. Simple algebraic operators of Eq.~(\ref{equationofpolesforTLSgeneralweights}) lead to a quadratic equation in $z$: 
\begin{eqnarray}
\label{equationofpolesforTLSentropyproduction}
 z^{-2}-2b(\nu) z^{-1}+\frac{ r_+}{r_-}=0,
\end{eqnarray} 
where the coefficient is given by 
\begin{eqnarray}
	 b(\nu)=\frac{1}{\Omega^2r_-}
	 \left[\xi^3(\nu)+\xi(\nu)(4\mu^2-r_-r_+)\right]. \nonumber
 \end{eqnarray}
The two SCGFs-FPTS are the two roots of this equation, which are   \begin{eqnarray}
\label{SCGFofentropycurrent}	\psi_{\pm}(\nu)=-\ln\left[ b(\nu)\pm\sqrt{b^2(\nu)-\frac{r_+}{r_-} }\right],
\end{eqnarray}
respectively. Considering that both these functions are real, we observe that the range of $\nu$ is restricted. Furthermore, because these two rates $r_{\pm}$ satisfy the detailed balance condition, we can directly verify that these two  SCGFs-FPTS~(\ref{SCGFofentropycurrent}) satisfy the fluctuation theorem~\cite{Lebowitz1999}:  
\begin{eqnarray}
	\psi_{+}(\nu)=\beta\omega_0 -\psi_{-}(\nu).  
\end{eqnarray} 
We illustrate these four functions in Fig.~\ref{fig2} for a set of parameters.

\begin{figure}
\includegraphics[width=1\columnwidth]{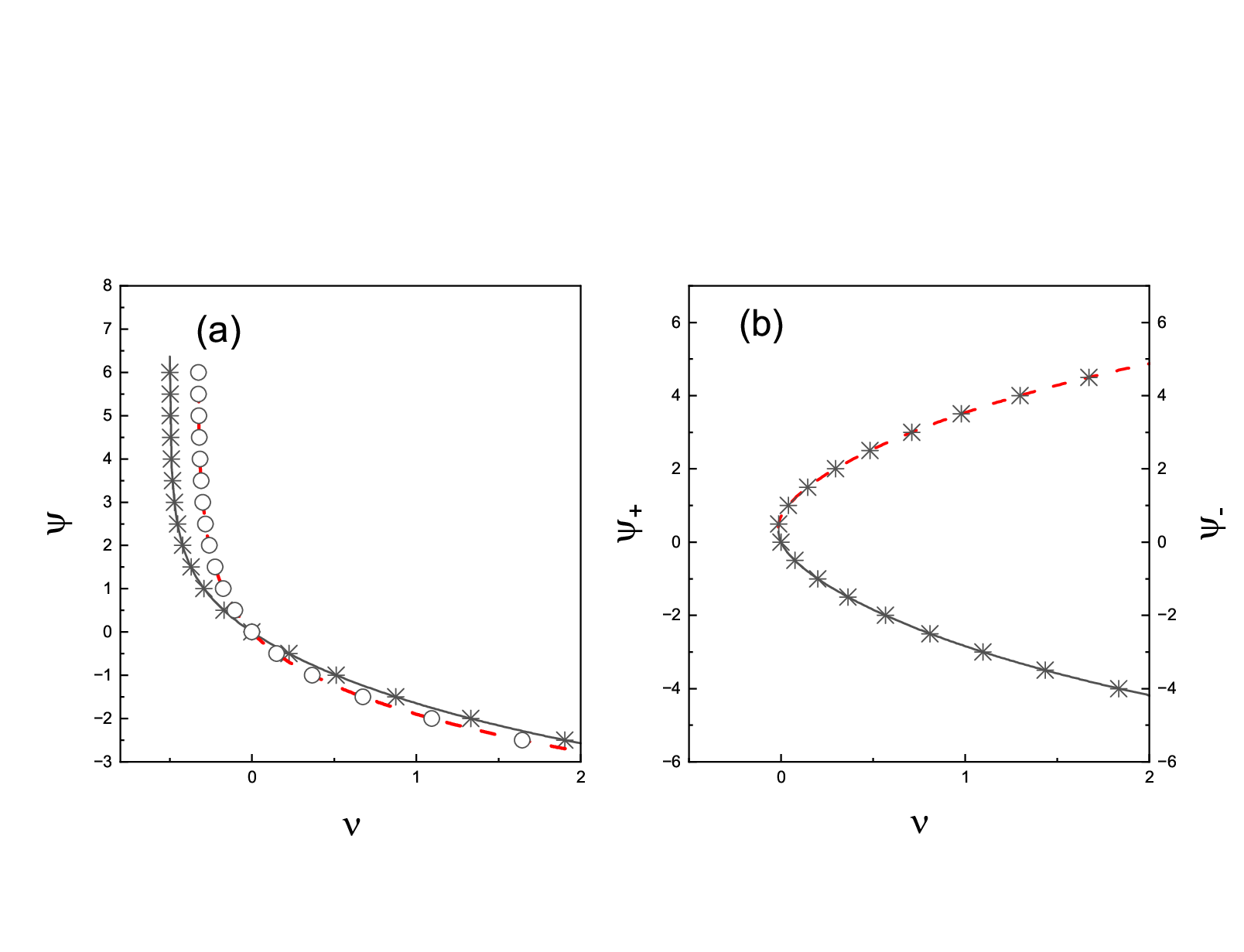}
\caption{{(a) The solid and dashed curves are the SCGFs-FPTS given by   Eqs.~(\ref{SCGFofTLScountingground}) and (\ref{SCGFofTLScounting12only}), respectively. Both counting variables are simple; the former counts the number of collapses to the ground state $|2\rangle$,  while the latter counts the number of the collapses where the TLS starts from $|1\rangle$ and ends in $|2\rangle$. (b) The solid and dashed curves are the SCGFs-FPTS given by Eq.~(\ref{SCGFofentropycurrent}) for the current-like variable. This variable is closely related to heat or entropy production in QJTs. In both panels, the physical parameters are set to $\omega_0=1$, $r_-=1$, $r_+=0.5$, and $\Omega=0.8$. Additionally, the star symbols and open circles indicate the SCGFs-FPTS solved for via the inverse function method for each counting variable.} }
\label{fig3}
\end{figure}

{To independently verify the data presented in Fig.~\ref{fig3}, we can employ the inverse function method to calculate the SCGFs-FPTS. To achieve this, we determine the SCGFs-CS for each counting variable by finding the largest real part of the roots of Eqs.~(\ref{equationofpolesforTLScountingground}), (\ref{equationofpolesforTLScounting12only}), and (\ref{equationofpolesforTLSentropyproduction}) in $\nu$ for a given $z=\exp(\lambda)$. The results are shown in Fig.~\ref{fig3}, indicated by the star symbols and open circles. In addition, we observe that the degrees of these polynomials in $\nu$ and $z$ differ: for the former, the degrees are $3$, $6$, and $3$, respectively, while for the latter, they are $1$, $1$, and $2$, respectively. It is a well-established fact that if the degree of a univariate polynomial equation is greater than or equal to 5, there are no general radical solutions to the equation~\cite{Speigel1968}. Consequently, we cannot explicitly write the SCGF-CS $\varphi(\lambda)$ as given by Eq.~(\ref{equationofpolesforTLScounting12only}). However, this does not imply that obtaining an analytical formula for the SCGF-FPTS $\psi(\nu)$ is impossible, as demonstrated by Eq.~(\ref{SCGFofTLScounting12only}). }

\subsection{Violations of classical uncertainty relations}
\label{section4b}
In the past several years, quantum violations of the TUR~\cite{Barato2015,Gingrich2016} and KUR~\cite{Garrahan2017} have attracted considerable attention.~\cite{Ptaszyifmmodenelsenfiski2018,Abah2014,Menczel2021,Kalaee2021,VanVu2022,Carollo2019,Hasegawa2020,Kewming2024}. Because both the uncertainty relations are conjugated in the CS and FPTS~\cite{Gingrich2017}, we are interested in applying the SCGFs-FPTS,  Eqs.~(\ref{SCGFofTLScountingground}) and~(\ref{SCGFofentropycurrent}), to study the same question. 

\subsubsection{Classical kinetic uncertainty relation}
The first is the classical KUR~\cite{Garrahan2007}. Its conjugation in the FPTS, expressed in terms of the SCGF, is  
\begin{eqnarray}
\label{FPTKURinSCGF}
-\left.\frac{\psi''}{\psi'}\right|_{\nu=0}\ge \frac{1}{\langle k\rangle},
\end{eqnarray}  	
where $\langle k\rangle$ represents the mean activity and the prime ($'$) and double prime ($''$) denote the first and second derivatives with respect to $\nu$, respectively. Here, the counting variable of interest is the number of collapses to the ground state $|2\rangle$, which was discussed in Sec.~(\ref{section4A}). Using Eq.~(\ref{SCGFofTLScountingground}), we can easily calculate the left-hand side of Eq.~(\ref{FPTKURinSCGF}), and the result is 
\begin{eqnarray}
\label{KURLHS}
\frac{2\Omega^2+r^2+4r_+r_-}{r_-(\Omega^2+rr_+)}-\frac{6r}{2\Omega^2+r^2}.
\end{eqnarray}

The mean activity on the right-hand side of Eq.~(\ref{FPTKURinSCGF}) is equal to the reciprocal of the average of the random variable ${\cal T}/n$ in the large $n$ limit, where $n$ is an integer and is a value of the dynamical activity $K$. We let the SCGF-FPTS for the counting variable $K$ be $\psi_k(\nu)$. The corresponding equation of poles is obtained by setting $w_{\alpha\beta}=1$, for $\alpha,\beta=1,2$, in Eq.~(\ref{equationofpolesforTLSgeneralweights}). Simple algebraic operations yield   
\begin{eqnarray}
	\label{equationofpolesforfullcountingvaraible}
2r_+r_-\xi(\nu) z^{-2} + r\Omega^2  z^{-1} -2\xi(\nu)(\xi^2(\nu)+4\mu^2)=0. 
\end{eqnarray}
{Interestingly, although the dynamical activity is a simple counting variable, Eq.~(\ref{equationofpolesforfullcountingvaraible}) is quadratic in $z$. According to Eq.~(\ref{SCGFdefinitionFPTtype1}), the SCGF-FPTS is the largest real root, which is given by 
\begin{eqnarray}
	\psi_k(\nu)=\ln\frac{4r_+r_-\xi(\nu)}{\sqrt{r^2\Omega^4+16r_+r_-\xi^2(\nu)[\xi^2(\nu)+4\mu^2]}-r\Omega^2 }. 
\end{eqnarray}}
Then, the right-hand side of Eq.~(\ref{FPTKURinSCGF}), or the classical kinetic bound, is 
\begin{eqnarray}
	\label{KURRHS} 
\frac{1}{\langle k\rangle}=-\psi_k'(0)=\frac{2\Omega^2+r^2}{r(\Omega^2+2r_+r_-)}. 
\end{eqnarray}

In Fig.~\ref{fig4}(a), using two independent dimensionless parameters, $r_0/\Omega$ and $\bar{n}$, we numerically depict the exact curve on which the inequality~(\ref{FPTKURinSCGF}) reduces to an equality. We observe that quantum violations of the KUR are not uncommon in the parameter space, as indicated by the shaded area. Equations~(\ref{KURLHS}) and~(\ref{KURRHS}) explain the validity of the KUR at very small and large values of $r_0/\Omega$. If the Rabi frequency is very large, $\it i.e.$, $r_0/\Omega\ll 1$, the two equations approximate $1/r_-$ and $1/r$, respectively. Clearly, $1/r_- >1/r$. If the Rabi frequency is very small, $\it i.e.$, $r_0/\Omega\gg 1$, the two equations are approximately the equations for the classical KUR in an incoherent TLS. Under these conditions, the inequality must hold~\cite{Garrahan2017}, as also demonstrated by algebraic analysis. Notably, these two cases are independent of $\bar n$. We find that quantum violations of the KUR mainly occur at smaller $\bar n$ value and intermediate Rabi frequencies. To illustrate this observation, in Fig.~\ref{fig4}(b), we specifically depict Eqs.~(\ref{KURLHS}) and~(\ref{KURRHS}) as functions of $r_0/\Omega$ at fixed $\bar n=0.15$. 

There are two quantum bounds for the KUR in the literature. On the basis of the SMP perspective, Carollo {\it et al.} derived a factor $\chi$ to replace the mean activity~\cite{Carollo2019}, \begin{eqnarray}
	\label{Carollochifactor}
\chi=\sum_{\alpha,\beta=1}^2 \frac{\sigma^2_{\beta\alpha}}{\tau^2_{\beta\alpha}  }R_{\beta\alpha}c_\alpha. 
\end{eqnarray}
The precise meanings of the involved coefficients are detailed in Appendix B. We present the numerical data for the quantum bound $1/\chi$ in Fig.~\ref{fig4}(b); see the dashed curve. Another quantum bound was proposed by Tan and Saito~\cite{VanVu2022}. They corrected the mean activity to 
\begin{eqnarray}
\label{TanSaitobound}
\langle k \rangle+Q=\langle k\rangle \left(1+ \frac{8\Omega^2}{r^2}\right).	
\end{eqnarray}
Because Refs.~\cite{VanVu2022} and \cite{Kewming2024} have provided careful derivations, we do not explain the correction further but directly present the final expression. We depict the quantum bound $1/(\langle k\rangle+Q)$ in Fig.~\ref{fig4}(b); see the dotted curve. We observe that these two quantum bounds are always valid. In particular, they uniformly converge to the classical bound at larger $r_0/\Omega$ values, which corresponds to that of the incoherent TLS. At larger Rabi frequencies, however, their predictions diverge significantly. Clearly, the bound of Eq.~(\ref{Carollochifactor}) is superior to that of Eq. (\ref{TanSaitobound}).

\subsubsection{Classical thermodynamic uncertainty relation}
\begin{figure}
	\includegraphics[width=1\columnwidth]{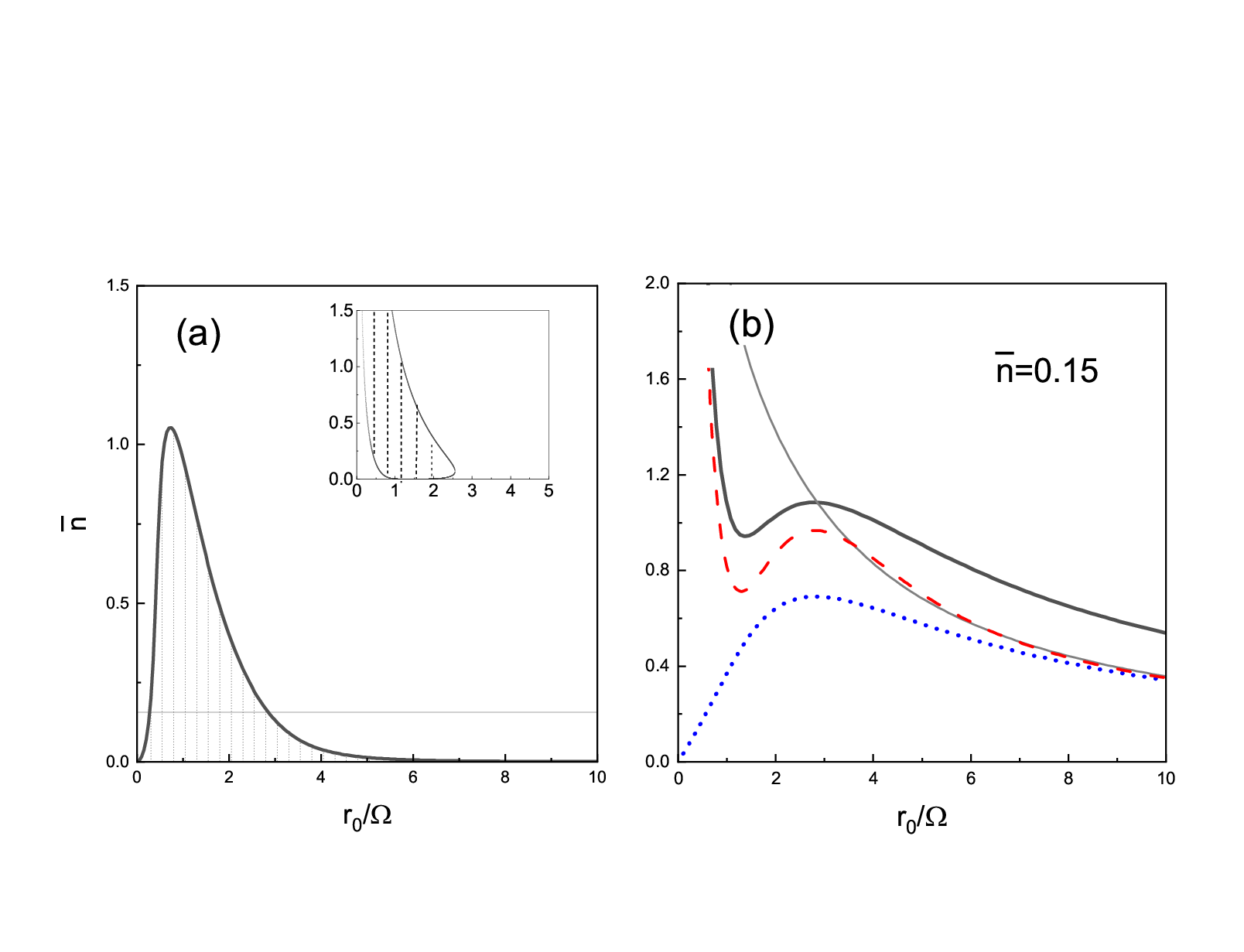}
	\caption{(a) The curve indicates the values of $r_0/\Omega$ ($x$-axis) and $\bar n$ ($y$-axis) at which the classical KUR for the quantum TLS reduces to an equality. The uncertainty relation is violated in the shaded area bounded by the curve and $x$-coordinate. The horizontal line indicates ${\bar n}=0.15$. The inset shows the curve on which the classical TUR for the same quantum system reduces to an equality. (b) The black solid curve represents the data calculated via the left-hand side of the classical KUR, Eqs.~(\ref{KURLHS}), while the gray thin curve represents the right-hand side (the classical bound), Eq.~(\ref{KURRHS}). The dashed and dotted curves are the quantum bounds $1/\chi$ and $1/(\langle k\rangle+Q)$, respectively.   }
	\label{fig4}
\end{figure}

The classical TUR for an asymmetric current in the FPTS, expressed in terms of the SCGFs,  is~\cite{Gingrich2017} 
\begin{eqnarray}
	\label{FPTTURinSCGF}
	-\left.\frac{\psi_+''}{\psi_+'}\right|_{\nu=0}\ge \frac{2k_B}{\langle \sigma\rangle},
\end{eqnarray} 
where $k_B$ is the Boltzmann constant, $\langle\sigma \rangle$ represents the mean entropy production rate, and we assume that the mean current is positive. {Although the current-like variable mentioned in Sec.~(\ref{section4A}) is not strictly asymmetric, we consider it to be a certain ``current" in the quantum regime and apply it to the classical TUR. The physical rationale is that this variable indeed reduces to a classic current when the quantum TLS becomes incoherent.} In this scenario, the general Eq.~(\ref{FPTTURinSCGF}) is simplified to 
\begin{eqnarray}
\label{FPTTURinSCGFentropyproduction}
\beta\omega_0\left.\frac{\psi''_{+}}{(\psi'_{+})^2}\right|_{\nu=0}\ge 2.
\end{eqnarray}
Here, we use the relation $k_B/\langle\sigma\rangle= -\psi_{+}'(0)/\beta\omega_0$. By substituting the exact Eq.~(\ref{SCGFofentropycurrent}) and performing simple algebraic operations, we obtain the left-hand side of Eq.~(\ref{FPTTURinSCGFentropyproduction}): \begin{eqnarray}
	\label{TURTLS}
\beta\omega_0\left[	\frac{r}{\delta}-\frac{6\Omega^2\delta r}{(2\Omega^2+r^2)^2}\right]. 
\end{eqnarray}
Similar to the KUR case, using the parameters $r_0/\Omega$ and $\bar n$, we numerically determine the curve on which the inequality~(\ref{FPTTURinSCGFentropyproduction}) reduces to an equality; see the inset of Fig.~\ref{fig4}(a). We observe that violations of the classical TUR are also not uncommon. In particular, the area of violation is distinct from that where the classical KUR is satisfied. Menczel {\it et al.}.~\cite{Menczel2021} investigated the causes of these violations in detail. They obtained the same Eq.~(\ref{TURTLS}) as a special case of a general equation in the presence of detuning; see Eq.~(10) therein. Note that the preceding derivations were carried out in the CS, where a characteristic polynomial technique was used~\cite{Bruderer2014}. Therefore, this consistency again demonstrates the conjugation relationship between the LDs of these two types of statistics.  

\section{Conclusions} \label{section5}
{In this work, we apply the SMP method to investigate the LDs-FPTS in a specific class of open quantum systems, where nonequilibrium quantum processes can be described by SMPs. Our results demonstrate that, consistent with the conjugation relationship found in classical stochastic systems and open quantum systems for simple counting variables, in the quantum regime, the LDs-FPTS and LDs-CS remain conjugated for the general counting variables. Specifically, the SCGFs-FPTS and SCGFs-CS are inverse functions of each other. 
		
During the demonstration of the conjugation relationship of the LDs, we present several calculation methods for the SCGFs-FPTS in these open quantum systems. For the simple counting variables, these functions are obtained by finding the largest modulus of the roots of the equation of poles with respect to the $z$-transform parameter $z$, and then calculating its logarithm. For the current-like variables, the root-based method generally fails, which is an unexpected observation. Interestingly, a practically important exception occurs when the equation of poles reduces to a quadratic equation in $z$. In such cases, the method is revived, and the SCGFs-FPTS can be solved for exactly. Otherwise, we resort to the inverse function method. These theoretical results and methods are demonstrated by a typical quantum TLS. Upon reflecting on these calculation methods for the SCGFs-FPTS, we perceive that their essence lies in determining the ROC for the joint transform in the simplest possible way. In fact, the condition of the spectral radius of the matrix $\mathbb W$ being less than $1$ can also be used to calculate these SCGFs. Compared with methods based on solving for the roots of the equation of poles, this method is more complicated and less efficient.  
 
In addition to the theoretical concerns about the SCGFs-FPTS, we also apply these functions to investigate quantum violations of the classical KUR and TUR. An intriguing finding is that the quantum bound based on the SMP is tighter than other quantum bounds. 
 
We conclude this work by discussing a limitation of studying the LDs of open quantum systems from the SMP perspective and an unsolved question arising from it. The applicability of the SMPs implies that the LDs of other open quantum systems where the SMP description is unsuitable are beyond the scope of our theory. Importantly, these systems are thought to be quite common~\cite{Landi2024}. On the other hand, the uniqueness of the SMPs is that, for the current-like variables, the crucial Saito-Dhar-Ptaszynski equation~(\ref{SatioDharequation}) is available in the specific open quantum systems. Conversely, this equation may not be applicable in other quantum systems. Thus, we encounter the question of whether the conjugation relationship between the LDs of the two types of statistics, particularly, the inverse function relationship between the SCGFs-FPTS and SCGs-CS, still holds. We hope to report this progress on this issue in the near future. }\\ \\
{\noindent\it Acknowledgments}
This work was supported by the National Natural Science Foundation of China under Grants No. 12075016 and 11575016.\\

\appendix

\section{Boundary equation~(\ref{equationofpoles})}
{We consider real values of $\nu$ and $z$. The convergence of the Neumann series for the matrix $\mathbb W$ is equivalent to the condition that the spectral radius of the matrix is less than $1$, that is, 
\begin{eqnarray}
	\varrho(\mathbb{W}(\nu,z))=\max_{i}{|\Lambda_i(\nu,z)|}<1,
\end{eqnarray} 
where $\Lambda_i$, for $i=1,\cdots, M$, are the eigenvalues of the matrix (see, {\it e.g.}, p618, Ref.~\cite{Meyer2023}). The modulus $|\Lambda_i(\nu,z)|$ of each eigenvalue is clearly less than 1. We express the determinant given in Eq.~(\ref{equationofpoles}) as a product of eigenvalues and demonstrate that   
\begin{eqnarray}
	\label{determinantproductionofeigens}
	\det[{\mathbb I}-{\mathbb W}(\nu,z)]=\prod_{i=1}^M (1-
	\Lambda_i)>0. 
\end{eqnarray}
In the product, if an eigenvalue $\Lambda_i$ is real, then $1-\Lambda_i>0$; otherwise, if an eigenvalue $\Lambda_i$ is a complex number, it must occur in a conjugate pair in a matrix with real elements. Thus, each pair $(\Lambda_i,\overline{\Lambda}_i)$ in this product results in  $(1-\Lambda_i)(1-\overline{\Lambda_i})=|1-\Lambda_i|^2>0$. Therefore, within the ROC, Eq.~(\ref{determinantproductionofeigens}), which is also the right-hand side of 
Eq.~(\ref{equationofpoles}), is strictly positive. Furthermore, if we consider the determinant as a continuous function of $\mu$ and $z$, points on the boundary of the ROC must satisfy the equation of poles. }
  
\section{Relationship between matrices $\mathbb G$ and $\mathbb W$  }
In the preceding work, we defined an $M\times M$ matrix $\mathbb G$ to calculate the MGF-CS of the specific class of open quantum systems: 
\begin{eqnarray}
	\label{countinggeneratingfunction}
	\hat Z_{C}(\lambda|\nu)&=&{\bf 1}^T {\mathbb G}^{-1}(z,\nu) {\bf P}(0).
\end{eqnarray} 
where the elements of the matrix $\mathbb G$ are~\cite{Liu2022} 
\begin{eqnarray}
	[{\mathbb G}]_{\gamma\gamma}&=&\frac{1-\hat p_{\gamma|\gamma}(\nu)z^{-w_{\gamma\gamma}}}{\hat S_\gamma(\nu)} \\ {\rm }
	[{\mathbb G}]_{\gamma\eta}&=&-\frac{\hat p_{\gamma|\eta}(\nu)z^{-w_{\gamma\eta}}}{\hat S_\gamma(\nu)}, {\hspace{0.1cm}  } \gamma\neq\eta.
\end{eqnarray}
It is easy to see that  
\begin{eqnarray}
	\label{Gmatrixcountingstatistics}
{\mathbb G}(\nu,z)=\left[\mathbb{ I - W}(\nu,z)\right]{\hat{\mathbb S}}^{-1}(\nu)
=\bar{\hat{{\mathbb T} }}^{-1}(z|\nu).  
\end{eqnarray}
Actually, there is still another expression:   
\begin{eqnarray}
	{\mathbb G}(z,\nu)&=&\nu{\mathbb I}-{\mathbb L}(z,\nu)
\end{eqnarray}
where the elements of the $M\times M$ matrix $\mathbb L$ are \begin{eqnarray}
	[{\mathbb L}]_{\gamma\gamma}&=&\frac{\hat p_{\gamma|\gamma}(\nu)z^{-\omega_{\gamma\gamma}}-\sum_{\alpha=1}^M {\hat p}_{\alpha|\gamma}(\nu)}{\hat S_\gamma(\nu)} 
\end{eqnarray}
and $[{\mathbb L}]_{\gamma\eta}=[{\mathbb G}]_{\gamma\eta}$ with $\gamma\neq\eta$. The matrix $\mathbb L$ is the quantum extension of the tilted generator of a classical SMP~\cite{Esposito2007b,Liu2022}. If all the weights $w_{\gamma\eta}$ and diagonal elements ${\hat p}_{\gamma|\gamma}$ are zero, it reduces to the standard generator of the SMP in the complex frequency domain~\cite{Qian2006}.

\section{The factor $\chi$}
In Eq.~(\ref{Carollochifactor}), the coefficients $c_\alpha$ for $\alpha=1,2$,  represent the rates of the quantum system collapsing in state $\phi_\alpha$ in the steady state. They can be written in terms of the Laplace transforms of the WTDs as~\cite{Liu2022,Carollo2019}   
\begin{eqnarray}
	c_1&=&\left.\frac{{\hat p}_{1|2}}{\tau_{1}{\hat p}_{1|2}+\tau_{2}{\hat p}_{2|1}}\right|_{\nu=0},\\
	c_2&=&\left.\frac{{\hat p}_{2|1}}{\tau_{1}{\hat p}_{1|2}+\tau_{2}{\hat p}_{2|1}}\right|_{\nu=0},
\end{eqnarray}
where $\tau_\alpha={\hat p}'_{1|\alpha}+{\hat p}'_{2|\alpha}$ are the mean times of the quantum system starting from state $\phi_\alpha$ and continuously evolving without collapse, that is, the mean ages of deterministic evolution. The coefficients $R_{\beta\alpha}$, for $\alpha,\beta=1,2$, are the transition probabilities of a Markov chain embedded in the SMP, which are simply equal to $\left.\hat p_{\beta|\alpha}\right|_{\nu=0}$. The coefficients $\sigma^2_{\beta\alpha}$ and $\tau_{\beta\alpha}$ are the variance and mean, respectively, of the time that the quantum system starts from the quantum state $\phi_\alpha$ and finally collapses to state $\phi_\beta$~\cite{Carollo2019}. Their ratio is \begin{eqnarray}
	\frac{\sigma^2_{\beta\alpha}}{\tau^2_{\beta\alpha}}=\left.\frac{{\hat p}''_{\beta|\alpha}{\hat p}_{\beta|\alpha}}{\hat{p}'^2_{\beta|\alpha}}\right |_{\nu=0}-1. 
\end{eqnarray}
By substituting the exact formulas for the Laplace transforms of the WTDs of the quantum TLS, we can obtain exact expressions for each term in Eq.~(\ref{Carollochifactor}). Because this is a straightforward procedure and several expressions are lengthy, we do not present them here. 

To make this work self-contained, the Laplace transforms of the WTDs of the quantum TLS are listed below: 
\begin{eqnarray}
\hat p_{1|1}(\nu)&=&r_+\frac{\Omega^2}{2\xi(\nu)[\xi^2(\nu)+4\mu^2] },\nonumber\\
\hat p_{1|2}(\nu)&=&r_+\frac{\xi^2(\nu)+\delta\xi(\nu)/2+\Omega^2/2}{\xi(\nu)[\xi^2(\nu)+4\mu^2]} \nonumber\\
\hat p_{2|1}(\nu)&=&r_-\frac{\xi^2(\nu)-\delta\xi(\nu)/2+\Omega^2/2}{\xi(\nu)[\xi^2(\nu)+4\mu^2]},\nonumber \\
\hat p_{2|2}(\nu)&=&r_-\frac{\Omega^2}{2\xi(\nu)[\xi^2(\nu)+4\mu^2] }.
\end{eqnarray}


\end{document}